\documentclass[twocolumn]{aastex702}
\usepackage{amsmath,amssymb,bm}
\usepackage{graphicx}

\shorttitle{QPEs from Magnetized sBH-Disk Transits}
\shortauthors{Wang et al.}

\begin{document}
	\raggedbottom

	\title{Quasi-periodic Eruptions from Recurrent Satellite Black Hole Transits through Magnetized Galactic Nucleus Accretion Disks}

	\author[orcid=0009-0006-4243-7755,gname=Na,sname=Wang]{Na Wang}
	\affiliation{Department of Astronomy, Xiamen University, Xiamen, Fujian 361005, China}
	\email{wangna9936@163.com}

	\author[orcid=0009-0008-9726-9431,gname=Jing-Tong,sname=Xing]{Jing-Tong Xing}
	\affiliation{Department of Astronomy, Xiamen University, Xiamen, Fujian 361005, China}
	\email{kriseme@163.com}

	\author[orcid=0000-0001-8678-6291,gname=Tong,sname=Liu]{Tong Liu}
	\affiliation{Department of Astronomy, Xiamen University, Xiamen, Fujian 361005, China}
    \affiliation{SHAO-XMU Joint Center for Astrophysics, Xiamen University, Xiamen, Fujian 361005, China}
	\email[show]{tongliu@xmu.edu.cn}
	\correspondingauthor{Tong Liu}

	\author[orcid=0000-0002-7329-9344,gname=Ya-Ping,sname=Li]{Ya-Ping Li}
	\affiliation{Shanghai Astronomical Observatory, Chinese Academy of Sciences, 80 Nandan Road, Shanghai 200030, China}
    \affiliation{SHAO-XMU Joint Center for Astrophysics, Xiamen University, Xiamen, Fujian 361005, China}
	\email{liyp@shao.ac.cn}

	\author[orcid=0000-0002-7299-4513,gname=Shuang-Liang,sname=Li]{Shuang-Liang Li}
	\affiliation{Shanghai Astronomical Observatory, Chinese Academy of Sciences, 80 Nandan Road, Shanghai 200030, China}
    \affiliation{SHAO-XMU Joint Center for Astrophysics, Xiamen University, Xiamen, Fujian 361005, China}
	\email{lisl@shao.ac.cn}

	\begin{abstract}
Quasi-periodic eruptions (QPEs) are recurrent soft X-ray flares from galactic nuclei, but their origin remains uncertain. The delayed ultraviolet (UV) counterpart detected in ZTF19acnskyy provides a new constraint on viable models. We present a two-channel model in which a satellite black hole (sBH) repeatedly crosses a nuclear accretion disk threaded by a large-scale magnetic field. Gravitational focusing and dynamical drag generate hot, optically thick ejecta whose expansion and photon diffusion power the soft X-ray QPE. For fiducial Bondi-scale parameters, the model yields a characteristic X-ray duration of $\sim10^3\ \mathrm{s}$ and luminosity of $\sim10^{42}\ \mathrm{erg\,s^{-1}}$. For ZTF19acnskyy, the model reproduces the observed day-scale X-ray duration and energetics. Simultaneously, the sBH motion compresses and bends the background magnetic field, triggering in-disk reconnection. The reconnection channel provides the energy budget and photon-diffusion delay required for the variable UV component. Unfavorable magnetic fields or diffusion times longer than the QPE recurrence period can weaken or smear out the UV signal, potentially explaining the lack of clear UV counterparts in other QPE sources.
	\end{abstract}


\section{Introduction}\label{sec:intro}

Quasi-periodic eruptions (QPEs) are recurrent soft X-ray outbursts from galactic nuclei hosting accreting massive black holes. They were first recognized in GSN 069, and more than ten QPE sources have now been identified, including RX J1301.9+2747, systems discovered by the extended ROentgen Survey with an Imaging Telescope Array (eROSITA), sources associated with tidal disruption events (TDEs), and candidates identified in archival X-ray observations \citep{2019Natur.573..381M,2020A&A...636L...2G,2021Natur.592..704A,2024A&A...684A..64A,2025ApJ...989...13A,2021ApJ...921L..40C,2024Natur.634..804N,2025ApJ...983L..39C,2026A&A...706L..15B}. During an eruption, the characteristic blackbody temperature commonly rises from several tens of electron volts in quiescence to $\gtrsim 100\ \mathrm{eV}$, while the peak soft X-ray luminosity reaches approximately $10^{41}$--$10^{43}\ \mathrm{erg\,s^{-1}}$. Individual eruptions last from less than an hour to approximately a day, with recurrence intervals ranging from hours to more than ten days. Some systems also show alternating long and short intervals, unequal successive eruptions, and substantial secular evolution \citep{2022A&A...662A..49A,2024A&A...690A..80A,2024ApJ...965...12C,2024A&A...692A..15G,2025NatAs...9..895H}. In several sources, the long and short intervals evolve while their sum remains comparatively stable, consistent with two events governed by a common orbital clock \citep{2024PhRvD.109j3031Z,2026ApJ..1006L..50L}. A viable model should therefore account for the soft spectrum, large radiative output, eruption duration, recurrence clock, and observed timing diversity.

Models proposed for QPEs include accretion-flow instabilities, unstable mass transfer, and repeated interactions between an orbiting secondary and a compact accretion disk \citep{2022ApJ...928L..18P,2023ApJ...952...32P,2025ApJ...989..196P,2023A&A...674L...1M,2024A&A...692A..15G,2025PASA...42e.130G}. In star--disk impact models, the stellar surface intercepts disk gas and generates optically thick ejecta whose expansion and radiative diffusion can produce a soft X-ray eruption \citep{2023ApJ...957...34L,2023A&A...675A.100F,2025ApJ...983...40V}. Asymmetric forward and backward ejecta may produce alternating bright and faint eruptions, although the contrast in nearly perpendicular impacts can become large enough to suppress the fainter event \citep{2026ApJ..1006L..50L}. Although a satellite black hole (sBH) lacks a solid surface, gravitational focusing and gaseous drag can heat disk gas and transfer orbital energy to the disk, while the transit timing is determined by the orbital and disk geometry \citep{2024PhRvD.109j3031Z}. 

Until recently, recurrent multiwavelength counterparts had not been securely established for QPEs. ZTF19acnskyy, hosted by the galaxy SDSS J133519.91+072807.4 (hereafter Ansky), was first identified as a newly active nucleus with unusually long QPEs and tentative ultraviolet (UV) variability \citep{2024A&A...688A.157S,2025NatAs...9..895H}. Its supermassive black hole (SMBH) mass is estimated to lie in the range $\log(M_\bullet/M_\odot)\sim 6-7$ \citep{2026ApJ..1000L..57G}. Its recurrence interval increased from approximately 4.5 days in 2024 to $\sim 10-12$ days in 2025 and $\sim14$ days by early 2026, while the X-ray individual burst durations increased from $\sim0.6$ to $\sim1.5$ days \citep{2025A&A...703A.263H,2026ApJ..1001L...6C}. Individual eruptions reach peak soft X-ray luminosities of $L_{\rm X,peak}\simeq4\times10^{43}\ {\rm erg\,s^{-1}}$ and radiate $\sim10^{48}$ erg in X-rays per eruption \citep{2026ApJ..1001L...6C}. High-cadence UV monitoring during the 2025 June--August campaign further revealed recurrent UV brightenings over five consecutive cycles, with directly observed flux variations of $\sim30\%$ for luminosity $\sim 5 \times 10^{41}~ \rm erg~s^{-1}$, and a delay of $0.96^{+0.38}_{-0.39}$ days relative to the X-ray eruptions \citep{2026ApJ..1000L..57G}. Shock-cooling models can produce UV-dominated eruptions under some disk conditions \citep{2024ApJ...963L...1L}, but the delayed UV signal also raises the possibility that part of the interaction energy is deposited within an optically thick layer and released only after radiative diffusion. Magnetic dissipation within a magnetized disk may provide such a delayed energy-release channel.

\begin{figure*}[t]
		\centering
		\includegraphics[width=0.6\textwidth]{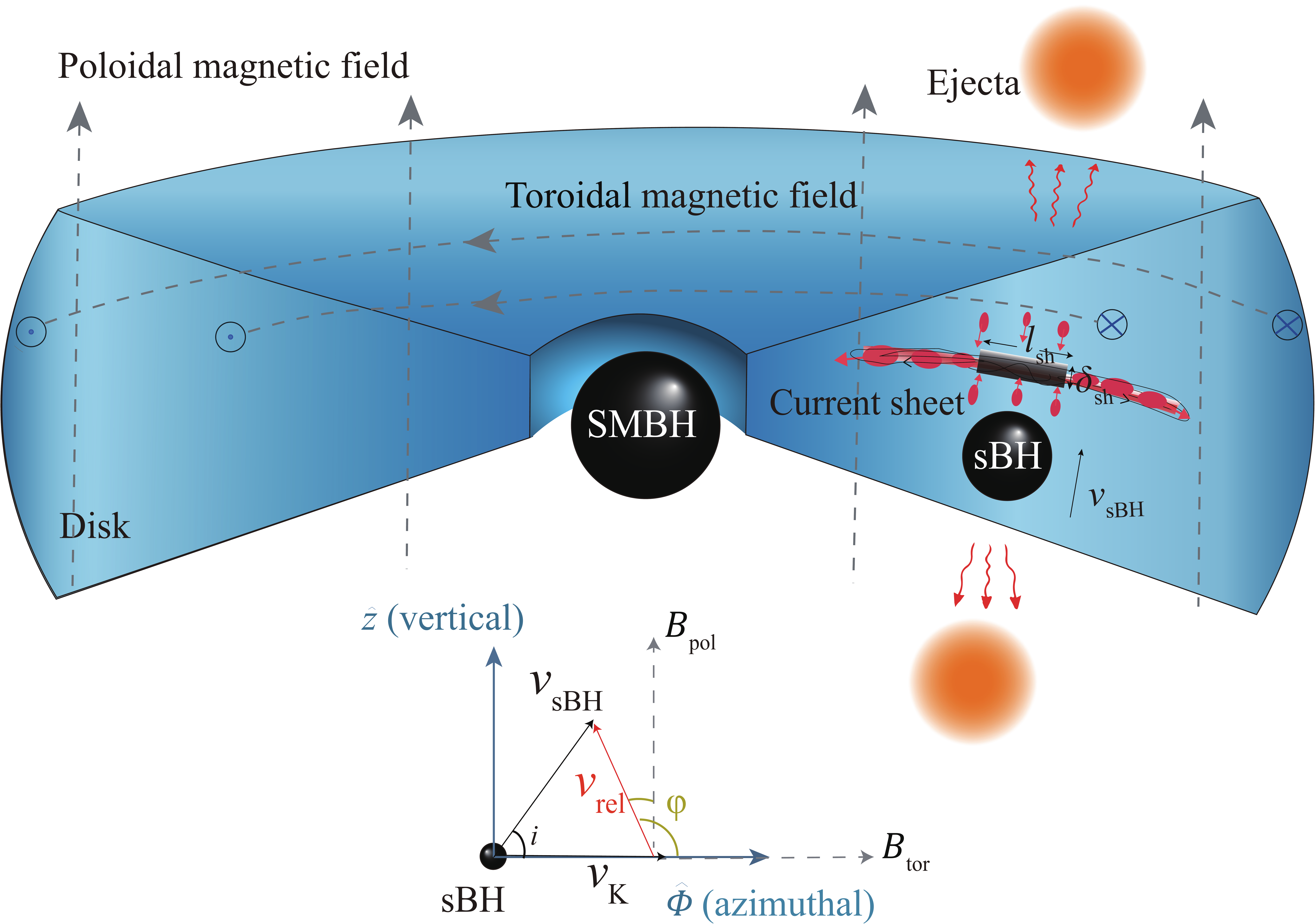}
		\caption{Schematic of recurrent satellite black hole (sBH) transits through a magnetized galactic nucleus accretion disk. An sBH on an orbit inclined by an angle $i$ repeatedly crosses the magnetized accretion disk surrounding a central SMBH. Gravitational focusing and dynamical drag heat the disk gas and lift optically thick material above and below the disk, producing expanding ejecta whose diffusion-powered emission produces the primary soft X-ray QPE. Simultaneously, the motion of the sBH relative to the disk gas compresses and bends the large-scale magnetic field, forming current sheets and triggering in-disk magnetic reconnection. The reconnection energy is thermalized at finite optical depth and diffuses toward the disk photosphere, where it can emerge as a broader and delayed UV counterpart. The lower inset illustrates the local velocity geometry at a disk-crossing node: the disk gas moves with the Keplerian velocity $v_{\mathrm K}\hat{\bm\phi}$, whereas the sBH moves with velocity $\bm v_{\mathrm{sBH}}$, giving $\bm v_{\mathrm{rel}}=\bm v_{\mathrm{sBH}}-v_{\mathrm K}\hat{\bm\phi}$. The poloidal and toroidal configurations represent two limiting orientations of the large-scale magnetic field, along $\hat{\bm z}$ and $\hat{\bm\phi}$, respectively. The schematic is not drawn to scale.}
		\label{fig1}
	\end{figure*}

In this $Letter$, we present a model for QPEs powered by recurrent sBH transits through a magnetized nuclear accretion disk. A single transit produces two radiative channels (Figure~\ref{fig1}): gravitational focusing and dynamical drag launch optically thick ejecta that power the soft X-ray QPE, while magnetic draping and reconnection deposit energy within the disk, with the associated radiation emerging later in the UV. We formulate the model framework in Section~\ref{sec:model}, derive the relevant scaling relations in Section~\ref{sec:radiation}, and discuss the observational implications and summarize our conclusions in Section~\ref{sec:discussion}.

\section{Physical Setup}
\label{sec:model}

\subsection{Disk and Orbital Geometry}
\label{subsec:disk_orbit}

We consider an sBH of mass $m_{\mathrm{sBH}}=10^3\,M_\odot$ orbiting a central SMBH of mass $M_\bullet=10^7\,M_\odot$, where $M_\odot$ denotes the solar mass. The numerical normalizations adopted below are reference scalings chosen to illustrate the broader QPE parameter space rather than source-specific fits. Ansky is treated separately as a source-specific benchmark in Section~\ref{sec:discussion}. The Schwarzschild radius of the central SMBH is $R_{\mathrm{s}}=2GM_\bullet/c^2$, and the collision radius is written as $R=rR_{\mathrm{s}}$, where $G$ is the gravitational constant, $c$ is the speed of light, and $r$ is the dimensionless collision radius. The local disk is approximated as geometrically thin, optically thick, and quasi-steady. Its vertical scale height is $H=hR$, where $h$ is the disk aspect ratio, and its Keplerian angular frequency is $\Omega_{\mathrm{K}}=(GM_\bullet/R^3)^{1/2}$. We adopt the $\alpha$-viscosity prescription, $\nu=\alpha H^2\Omega_{\mathrm K}$, where $\alpha$ is the dimensionless viscosity parameter. Neglecting the inner-boundary correction, the quasi-steady relation $\dot M\simeq3\pi\nu\Sigma$ gives \citep{1973A&A....24..337S,2023ApJ...957...34L}
\begin{equation}
	\Sigma \simeq 1.6\times10^4\,
	\dot m_{-1}\eta_{-1}^{-1}\alpha_{-2}^{-1}
	h_{0.05}^{-2}r_{50}^{-1/2}
	\ \mathrm{g\,cm^{-2}}.
	\label{eq:sigma}
\end{equation}
Here, the dimensionless accretion rate is $\dot m\equiv\dot M/\dot M_{\mathrm{Edd}}$, and $\dot M_{\mathrm{Edd}}=4\pi GM_\bullet/\eta\kappa c$ is the Eddington accretion rate, where $\eta$ is the radiative efficiency of accretion, and we adopt a constant electron-scattering opacity $\kappa=0.34\ \mathrm{cm^2\,g^{-1}}$ throughout this work. We define $r_{50}=r/50$, $\dot m_{-1}=\dot m/0.1$, $\eta_{-1}=\eta/0.1$, $\alpha_{-2}=\alpha/0.01$, and $h_{0.05}=h/0.05$.

Near a disk crossing, we approximate both the disk gas and the sBH as moving on locally circular Keplerian orbits and neglect their radial velocity components. The disk gas rotates with velocity $\bm v_{\mathrm{d}}=v_{\mathrm{K}}\hat{\bm\phi}$, where $v_{\mathrm{K}}=c(2r)^{-1/2}$ is the local Keplerian speed. Here $\hat{\bm\phi}$ and $\hat{\bm z}$ denote the azimuthal and vertical unit vectors, respectively. For a circular orbit inclined by an angle $i$ relative to the disk plane, the sBH velocity at a disk-crossing node is $\bm v_{\mathrm{sBH}}=v_{\mathrm{K}}\cos i\,\hat{\bm\phi}+v_{\mathrm{K}}\sin i\,\hat{\bm z}$. We take $0\leq i\leq\pi$, so that both prograde and retrograde configurations are included. The velocity of the sBH relative to the disk gas is $\bm v_{\mathrm{rel}}=\bm v_{\mathrm{sBH}}-\bm v_{\mathrm{d}}$, with magnitude $v_{\mathrm{rel}}=2v_{\mathrm{K}}\sin\left(\frac{i}{2}\right)$. The magnitude of its vertical component is $v_z=v_{\mathrm{K}}|\sin i|$. For a circular orbit, the sBH fully exits the disk between successive crossings if its maximum vertical displacement exceeds the disk scale height, i.e., $|\sin i|>h$.

\subsection{Background Magnetic Field}
\label{subsec:background_field}

Motivated by accretion-powered magnetic-field scalings commonly adopted for compact accretion flows \citep{1997MNRAS.292..887G,1999ASPC..190..173L,2017NewAR..79....1L,2025ApJ...991..167X}, we parameterize the unperturbed large-scale magnetic field as
\begin{equation}
	\begin{aligned}
		B_0(R)
		&=\left(\frac{2\dot M c}{R^2}\right)^{1/2}\\
		&\simeq 2.1\times10^3\,
		\dot m_{-1}^{1/2}\eta_{-1}^{-1/2}
		M_7^{-1/2}r_{50}^{-1}\ \mathrm{G},
	\end{aligned}
	\label{eq:B0}
\end{equation}
where $M_7=M_\bullet/(10^7\,M_\odot)$. This prescription represents a dynamically subdominant background field rather than a self-consistent global solution for a magnetized accretion disk.

The magnetic pressure of the unperturbed field is $P_{\mathrm{B}_0}=B_0^2/(8\pi)$. We estimate the characteristic midplane pressure as $P_{\mathrm{mid}}\simeq\rho_{\mathrm{mid}}c_{\mathrm{s}}^2$, where the midplane mass density $\rho_{\mathrm{mid}}\simeq\Sigma/(2H)$ and the sound speed $c_{\mathrm{s}}\simeq H\Omega_{\mathrm{K}}$. We find $P_{\mathrm{B}_0}/P_{\mathrm{mid}}\simeq7.5\times10^{-3}$ for the fiducial parameters. Consequently, for the fiducial parameters adopted in this work, the magnetic pressure of the uncompressed background field is much lower than the characteristic pressure at the disk midplane. The background field therefore does not provide the dominant vertical pressure support near the midplane, consistent with treating it as dynamically subdominant in the adopted thin-disk model. We consider two limiting magnetic-field geometries: a poloidal field perpendicular to the disk plane, $\bm{B}_{\mathrm{pol}}=B_0\hat{\bm{z}}$, and a toroidal field along the local azimuthal direction, $\bm{B}_{\mathrm{tor}}=B_0\hat{\bm{\phi}}$. We define $\varphi$ as the acute angle between $\bm v_{\mathrm{rel}}$ and the poloidal-field direction. Only the component of the relative velocity perpendicular to the background magnetic field directly compresses and bends the field lines. We therefore define $v_{\perp B} \equiv\left|\bm{v}_{\mathrm{rel}}\times\hat{\bm{B}}\right|$, where $\hat{\bm{B}}=\bm{B}/B_0$ is the unit vector along the unperturbed magnetic field. The components of $\bm v_{\mathrm{rel}}$ perpendicular to the two field geometries are therefore
\begin{equation}
	v_{\perp B,\mathrm{pol}}
	=2v_{\mathrm{K}}\sin^2\left(\dfrac{i}{2}\right)
\end{equation}
	and
\begin{equation}
	v_{\perp B,\mathrm{tor}}
	=v_{\mathrm{K}}|\sin i|
	\label{eq:vperp}
\end{equation}
for the poloidal and toroidal magnetic fields, respectively. The resulting in-disk magnetic reconnection and emission are discussed in Section~\ref{subsec:reconnection}.

\section{Radiation from sBH--Disk Transits}
\label{sec:radiation}

\subsection{Soft X-ray emission}
\label{subsec:xray}

An sBH has no solid surface, so its effective interaction cross section is set by gravitational focusing \citep{1944MNRAS.104..273B,1952MNRAS.112..195B}. We characterize this scale using the Bondi radius $R_{\mathrm{B}}=2Gm_{\mathrm{sBH}}/v_{\mathrm{rel}}^2$. For the orbital geometry adopted here, the gas column density encountered by the sBH along the direction of relative motion is $\Sigma_{\mathrm{eff}}=\Sigma/\cos(i/2)$. The characteristic mass of gas heated and lifted out of the disk is therefore $M_{\mathrm{ej}}\simeq\pi R_{\mathrm{B}}^2\Sigma_{\mathrm{eff}}$. The numerical simulations of \citet{2026ApJ..1006L..50L} show that the region substantially shock heated and gravitationally perturbed by an sBH can extend beyond the nominal Bondi radius. The above expression should therefore be regarded as a conservative Bondi-scale estimate of the gas mass participating in the radiative process.

As the shock-heated gas expands with velocity $v_{\mathrm{ej}}$, its radius is $R_{\mathrm{ej}}\simeq v_{\mathrm{ej}}t$, and its radial optical depth is approximately $\tau(t)\simeq\kappa M_{\mathrm{ej}}/(4\pi R_{\mathrm{ej}}^2)$. The photon-diffusion timescale is $t_{\mathrm{diff,ej}}\sim\tau R_{\mathrm{ej}}/c$. Radiation begins to escape efficiently when $t_{\mathrm{diff,ej}}\sim t$. Adopting $v_{\mathrm{ej}}\simeq v_{\mathrm{rel}}$, the characteristic photon-escape timescale is
\begin{equation}
	\begin{aligned}
		t_{\mathrm{ej}}
		&\simeq
		\left(
		\frac{\kappa M_{\mathrm{ej}}}
		{4\pi c v_{\mathrm{ej}}}
		\right)^{1/2}\\
		&\simeq
		2.1\times10^3\
		\dot m_{-1}^{1/2}\eta_{-1}^{-1/2}\\
		&\quad\times
		\alpha_{-2}^{-1/2}h_{0.05}^{-1}
		m_3r_{50}
		\left[
		\frac{\sin(i/2)}{\sin(\pi/20)}
		\right]^{-5/2}\\
		&\quad\times
		\left[
		\frac{\cos(i/2)}{\cos(\pi/20)}
		\right]^{-1/2}~\mathrm{s},
	\end{aligned}
	\label{eq:tej}
\end{equation}
where $m_3=m_{\mathrm{sBH}}/(10^3\,M_\odot)$. This Arnett-like estimate provides an order-of-magnitude estimate of the characteristic duration of the soft X-ray QPE \citep{1982ApJ...253..785A,2023ApJ...957...34L,2025ApJ...983...40V}.

In this picture, the gravitational field of the sBH focuses and deflects the surrounding disk gas, generating a shock structure and a downstream gravitational wake. The disturbed gas, in turn, exerts a gravitational drag force on the sBH opposite to its relative motion. This force does negative work on the sBH, reducing its orbital mechanical energy and transferring the corresponding energy to the disturbed disk gas. The deposited energy can be partitioned among the internal energy of the shocked gas, bulk kinetic energy, and disordered motions in the wake.

Only a fraction of the heated gas can leave the disk and form hot, optically thick, expanding ejecta. In the semianalytic treatment adopted here, $E_{\mathrm{dep}}$ therefore represents the total energy transferred by the sBH to all disturbed disk gas during a single disk crossing, whereas $M_{\mathrm{ej}}$ characterizes the gas mass forming the expanding radiating structure. The bulk kinetic energy of the ejecta, $E_{\mathrm{kin,ej}}\simeq M_{\mathrm{ej}}v_{\mathrm{ej}}^2/2$, represents only the energy associated with the ordered expansion of the Bondi-scale ejecta. \citet{2024PhRvD.109j3031Z} noted that the gas mass estimated from the accretion cross section accounts for only a fraction of the shocked material, whereas the actual shock-perturbed region can extend beyond the accretion radius. We therefore use $E_{\mathrm{dep}}$ as the total energy budget available for heating, acceleration, and radiation.

The gravitational drag force can be approximated as \citep{2016A&A...589A..10T,2024PhRvD.109j3031Z}
\begin{equation}
	\bm F_{\mathrm{drag}}
	\simeq
	4\pi\ln\Lambda\,
	\frac{G^2\rho_{\mathrm{mid}}m_{\mathrm{sBH}}^2}
	{v_{\mathrm{rel}}^3}
	\bm v_{\mathrm{rel}}.
\end{equation}
The Coulomb logarithm is defined as $\ln\Lambda=\ln(b_{\max}/b_{\min})$. For the disk-crossing geometry considered here, we adopt $b_{\max}\simeq2H/\cos(i/2)$ as the maximum extent of the gravitational wake along the crossing path and $b_{\min}\simeq R_{\mathrm{B}}/2$ as the stand-off distance of the bow shock \citep{2024PhRvD.109j3031Z}. We adopt $\ln\Lambda=5$ for the fiducial parameters. As an alternative inner-cutoff prescription, one may take $b_{\min}\simeq Gm_{\mathrm{sBH}}/c^2$, which gives $\ln\Lambda\simeq11.5$ for the fiducial parameters. Integrating the gravitational drag force along the disk-crossing path of the sBH gives
\begin{equation}
	E_{\mathrm{dep}}\simeq4\pi\ln\Lambda\,\frac{G^2m_{\mathrm{sBH}}^2\Sigma}{v_{\mathrm{rel}}^2\cos(i/2)}.
\end{equation}

We define $\epsilon_{\mathrm X}$ as the fraction of the deposited energy that ultimately emerges as observable soft X-ray radiation from the optically thick ejecta, accounting collectively for coupling to the ejecta, adiabatic losses, radiative conversion, and the soft X-ray band correction. In the diffusion-limited approximation, if the injection timescale does not exceed $t_{\mathrm{ej}}$, the characteristic soft X-ray luminosity can be estimated as $L_{\mathrm X}\simeq\epsilon_{\mathrm X}E_{\mathrm{dep}}/t_{\mathrm{ej}}$. We adopt $\epsilon_{\mathrm X}=0.1$ as a fiducial value corresponding to a relatively efficient conversion scenario. If energy injection persists for a longer time, the longer of the injection and diffusion timescales should instead set the effective radiative-release timescale. The expression below should therefore be interpreted as a characteristic luminosity in the diffusion-limited regime, rather than necessarily as the instantaneous peak luminosity of an observed QPE. For the fiducial Bondi-scale normalization, we obtain
\begin{equation}
	\begin{aligned}
		L_{\mathrm X}
		&\simeq
		9.6\times10^{41}\ 
		\epsilon_{\mathrm X,-1}
		\left(\frac{\ln\Lambda}{5}\right)
		\dot m_{-1}^{1/2}
		\eta_{-1}^{-1/2}
		\alpha_{-2}^{-1/2}\\
		&\quad\times
		h_{0.05}^{-1}m_3r_{50}^{-1/2}
		\left[
		\frac{\sin(i/2)}{\sin(\pi/20)}
		\right]^{1/2}\\
		&\quad\times
		\left[
		\frac{\cos(i/2)}{\cos(\pi/20)}
		\right]^{-1/2}~\mathrm{erg\,s^{-1}}.
	\end{aligned}
	\label{eq:LX}
\end{equation}
Here, $\epsilon_{\mathrm X,-1}=\epsilon_{\mathrm X}/0.1$. The characteristic duration and luminosity are therefore controlled primarily by the sBH mass, orbital inclination, disk column density, and radiative properties of the ejecta. The reference normalization adopted here is intended only to illustrate the parameter scalings and does not correspond to any particular QPE.

\subsection{In-disk reconnection and delayed UV emission}\label{subsec:reconnection}

The motion of an sBH relative to the disk gas compresses and bends the magnetic flux frozen into the plasma \citep{2008ApJ...677..993D,2025ApJ...991..167X}. We model the resulting magnetic-field bending and shear as generating current sheets and triggering magnetic reconnection \citep{2020ApJ...900..100R,2022MNRAS.513.4267N,2025ApJ...991..167X}. Only the component of the relative velocity perpendicular to the background magnetic field directly compresses and bends the field lines. The corresponding ram pressure is therefore $P_{\mathrm{ram}}=\rho_{\mathrm{mid}}v_{\perp B}^{2}$. When $P_{\mathrm{ram}}\gtrsim P_{\mathrm{B}_0}$, the transverse flow driven by the sBH can substantially compress and distort the unperturbed background field \citep{2008ApJ...677..993D}. For the fiducial parameters adopted in this work, both magnetic-field geometries satisfy this triggering condition, with the toroidal configuration experiencing the larger ram pressure.

Once this condition is satisfied, the plasma motion around the sBH can continuously accumulate magnetic flux toward the current-sheet region and amplify the local magnetic field, thereby providing magnetic energy that can subsequently be dissipated through reconnection. We define the magnetic pressure associated with the compressed field in the current sheet as $P_{\mathrm{mag}}=B_{\mathrm{m}}^{2}/(8\pi)$. Following the phenomenological prescription of \citet{2025ApJ...991..167X}, we define $\beta_{\mathrm p}\equiv P_{\mathrm{ram}}/P_{\mathrm{mag}}$ and adopt the fiducial value $\beta_{\mathrm p}=0.01$. The compressed magnetic field is then $B_{\mathrm{m}}=v_{\perp B}(8\pi\rho_{\mathrm{mid}}/\beta_{\mathrm p})^{1/2}$, and we define $\beta_{-2}=\beta_{\mathrm p}/10^{-2}$. This value is representative of low-$\beta_{\mathrm p}$ conditions in strongly magnetized environments, including active galactic nucleus (AGN) coronae, the bases of relativistic jets, and highly magnetized surface layers of accretion disks. Numerical simulations indicate that low-$\beta_{\mathrm p}$ conditions facilitate the thinning of current sheets to kinetic scales and the onset of fast magnetic reconnection \citep{2018ApJ...852...95N}. Motivated by the elongated current-sheet geometries found in relativistic reconnection simulations \citep{2021JPlPh..87e9012R,2022ApJ...924L..32R}, we adopt $l_{\mathrm{sh}}=20r_{\mathrm{s}}$ and $\delta_{\mathrm{sh}}=0.1l_{\mathrm{sh}}$, where $l_{\mathrm{sh}}$ and $\delta_{\mathrm{sh}}$ are the length and width of the current sheet, respectively, and $r_{\mathrm{s}}=2Gm_{\mathrm{sBH}}/{c^2}$ is the Schwarzschild radius of the sBH.

Before reconnection begins, the magnetic field is approximately frozen into the disk plasma. For simplicity, we adopt the full disk thickness, $2H$, as the effective magnetic-flux collection scale for both background-field geometries. During a time interval $t_{\mathrm{form}}$, the background magnetic flux swept into the current sheet by the sBH motion can be estimated as $\Phi_{\mathrm{in}}\sim2HB_0v_{\perp B}t_{\mathrm{form}}$. The compressed magnetic flux contained in the current sheet is approximately $\Phi_{\mathrm{sh}} \sim B_{\mathrm{m}}l_{\mathrm{sh}}\delta_{\mathrm{sh}}$. Because magnetic flux is conserved, the formation timescale is $t_{\mathrm{form}}\sim\frac{l_{\mathrm{sh}}\delta_{\mathrm{sh}}}{2HB_0}\frac{B_{\mathrm{m}}}{v_{\perp B}}$. The two magnetic-field geometries therefore have the same characteristic formation timescale.

We characterize the magnetization of the plasma upstream of the current sheet by $\sigma=\frac{B_{\mathrm{m}}^{2}}{4\pi\rho_{\mathrm{mid}}c^2}$, where the upstream density has been approximated as $\rho_{\mathrm{mid}}$. The corresponding relativistic Alfvén speed is $v_{\mathrm A}=c[\sigma/(1+\sigma)]^{1/2}$ \citep{2022A&A...663A.169E,2023A&A...677A..67E}. Because a low value of $\beta_{\mathrm p}$ does not necessarily imply $\sigma\gg1$, we retain the full relativistic expression for the Alfvén speed rather than adopting the limit $v_{\mathrm A}\simeq c$. We write the reconnection inflow speed as $v_{\mathrm{in}}=\mathcal R_{\mathrm{rec}}v_{\mathrm A}$, where $\mathcal R_{\mathrm{rec}}$ is the dimensionless reconnection rate. The reconnection timescale can then be estimated as the time required for the plasma to cross the characteristic current-sheet thickness at the reconnection inflow speed: $t_{\mathrm{rec}}\sim \frac{\delta_{\mathrm{sh}}}{\mathcal R_{\mathrm{rec}}v_{\mathrm A}}$, where $\mathcal R_{-1}=\mathcal R_{\mathrm{rec}}/0.1$. For the fiducial parameters, $t_{\mathrm{rec}}$ is of order a few seconds. The time required for the sBH to traverse the full disk thickness is
\begin{equation}
	t_{\mathrm{cross}}
	\simeq
	\frac{2H}{v_z}
	=
	1.6\times10^4\,
	h_{0.05}M_7r_{50}^{3/2}
	\left[
	\frac{|\sin i|}
	{\sin(\pi/10)}
	\right]^{-1}
	\ \mathrm{s}.
	\label{eq:tcross}
\end{equation}
Within the parameter regime in which magnetic compression can be triggered, one generally has $t_{\mathrm{form}},\quad t_{\mathrm{rec}}\ll t_{\mathrm{cross}}$. Magnetic-flux accumulation and local reconnection therefore occur repeatedly during a single disk crossing. These local timescales, which are of order seconds, do not directly determine the observed UV variability on hour-to-day timescales. Here $t_{\mathrm{rec}}$ characterizes the local dissipation time of an individual current sheet, rather than the duration of the observed UV emission. The characteristic duration of the UV response is therefore set by the longer of the macroscopic energy-injection and photon-diffusion timescales. The macroscopic energy injection is maintained by an ensemble of current sheets during $t_{\mathrm{cross}}$.  

Because reconnection occurs inside an optically thick accretion disk, the released energy cannot escape immediately. If the dissipation occurs near the disk midplane, the optical depth to one disk surface is approximately $\tau_\perp \simeq \frac{\kappa\Sigma}{2}$ \citep{1979rpa..book.....R}. The corresponding photon diffusion timescale is
\begin{equation}
	t_{\mathrm{diff}}
	\simeq
	\frac{\tau_\perp H}{c}
	\simeq
	6.6\times10^5\
	\dot m_{-1}\eta_{-1}^{-1}
	\alpha_{-2}^{-1}
	h_{0.05}^{-1}
	M_7r_{50}^{1/2}
	\ \mathrm{s}.
	\label{eq:tdiff}
\end{equation}
This estimate corresponds to a timescale of several days and is regarded as an upper limit for dissipation occurring deep within the disk. Gravitational focusing, shocks, and gas uplift induced by the sBH transit may redistribute the local disk gas and create a relatively low-density channel above the dissipation region, thereby reducing the remaining column density and the photon diffusion time.

We next estimate the magnetic dissipation power supplied by multiple current-sheet structures during a single disk crossing. The effective volume of a representative current sheet is written as $V_{\mathrm{sh}} \simeq \left(v_{\mathrm{rel}}t_{\mathrm{rec}}\right)l_{\mathrm{sh}}\delta_{\mathrm{sh}}$, where $v_{\mathrm{rel}}t_{\mathrm{rec}}$ represents the azimuthal orbital displacement during one local reconnection time. In the inclined disk-crossing configuration considered here, the current sheet is generated by the relative motion between the sBH and the disk gas. We therefore use $v_{\mathrm{rel}}t_{\mathrm{rec}}$ as the effective third dimension of the reconnection structure along the transit wake. The magnetic energy available in one representative current sheet is $E_{\mathrm{sh}} \simeq \frac{B_{\mathrm{m}}^{2}}{8\pi} V_{\mathrm{sh}}$. We interpret $E_{\mathrm{sh}}/t_{\mathrm{rec}}$ as the average dissipation power of a representative current sheet. During a disk crossing, current sheets are continuously generated and supplied with magnetic flux. The total thermal power is therefore estimated by
summing over the effective number of reconnecting structures,
\begin{equation}
	L_{\mathrm{th}}
	\simeq
	N_{\mathrm{sh}}
	\frac{E_{\mathrm{sh}}}
	{t_{\mathrm{rec}}}
	\simeq
	\frac{B_{\mathrm{m}}^{2}}{8\pi}
	\frac{t_{\mathrm{cross}}}
	{t_{\mathrm{form}}}
	v_{\mathrm{rel}}
	l_{\mathrm{sh}}\delta_{\mathrm{sh}},
	\label{eq:Lheat}
\end{equation}
where $N_{\mathrm{sh}}\simeq\frac{t_{\mathrm{cross}}}{t_{\mathrm{form}}}$ represents the effective multiplicity of current sheets accumulated and continuously maintained during one disk crossing. Thus, $t_{\mathrm{rec}}$ enters Equation~(\ref{eq:Lheat}) only through the local dissipation power of an individual current sheet, whereas $t_{\mathrm{cross}}$ sets the duration of the macroscopic energy injection.

Using the above expressions for $B_{\mathrm m}$ and $t_{\mathrm{form}}$ together with Equation~(\ref{eq:tcross}), we obtain the two background magnetic-field geometries yield
\begin{equation}
	\begin{aligned}
		L_{\mathrm{th,pol}}
		\simeq{}&
		6.9\times10^{41}\,
		\beta_{-2}^{-1/2}
		\dot m_{-1}\eta_{-1}^{-1}
		\alpha_{-2}^{-1/2}
		h_{0.05}^{1/2}
		M_7r_{50}^{-3/4}
		\\
		&
		{}\times
		\left[
		\frac{\sin(i/2)}
		{\sin(\pi/20)}
		\right]^4
		\left[
		\frac{\cos(i/2)}
		{\cos(\pi/20)}
		\right]^{-1}
		\ \mathrm{erg\,s^{-1}},
\end{aligned}
\end{equation}
and
\begin{equation}
\begin{aligned}
		L_{\mathrm{th,tor}}
		\simeq{}&
		2.7\times10^{43}\,
		\beta_{-2}^{-1/2}
		\dot m_{-1}\eta_{-1}^{-1}
		\alpha_{-2}^{-1/2}
		h_{0.05}^{1/2}
		M_7r_{50}^{-3/4}
		\\
		&
		{}\times
		\left[
		\frac{\sin(i/2)}
		{\sin(\pi/20)}
		\right]^2
		\left[
		\frac{\cos(i/2)}
		{\cos(\pi/20)}
		\right]
		\ \mathrm{erg\,s^{-1}}.
	\end{aligned}
	\label{eq:Lth}
\end{equation}
These estimates assume that magnetic flux is supplied continuously throughout $t_{\mathrm{cross}}$. Intermittent flux supply or limited temporal overlap among different reconnection structures would reduce the effective multiplicity and the total heating rate. For the standard thin disk adopted here, the local effective temperature is $T_{\mathrm{eff}}\simeq [3GM_\bullet\dot M/(8\pi\sigma_{\mathrm{SB}}R^3)]^{1/4} \simeq3.5\times10^4\, \dot m_{-1}^{1/4}\eta_{-1}^{-1/4} M_7^{-1/4}r_{50}^{-3/4}\ \mathrm{K}$, where $\sigma_{\mathrm{SB}}$ is the Stefan--Boltzmann constant \citep{1973A&A....24..337S}. At larger disk-crossing radii, the lower effective temperatures shift the reprocessed thermal emission toward the UV, allowing a substantial UV component. If the reconnection energy is efficiently thermalized and reprocessed by the optically thick disk, the resulting UV luminosity is comparable to $L_{\mathrm{th}}$ in order of magnitude.

A detectable UV counterpart first requires the in-disk reconnection heating to produce a reprocessed component that can be distinguished from the quiescent nuclear UV emission and exceeds the instrumental detection threshold. Within the present closure prescription, a weaker background field gives $t_{\mathrm{form}}\propto B_0^{-1}$, which slows magnetic-flux accumulation and reduces both the effective reconnection multiplicity and the total thermal luminosity. Conversely, if the background field is sufficiently strong that $P_{\mathrm{B}_0}\gtrsim P_{\mathrm{ram}}$, the transverse ram pressure driven by the sBH can no longer substantially compress and bend the field. Detectable UV emission is therefore favored within an intermediate range of background-field strengths: the disk must contain sufficient magnetic flux to provide appreciable reconnection power, while the unperturbed field must remain susceptible to perturbation by the sBH transit. The diffusion time primarily determines the temporal appearance of the UV response rather than whether the response is produced. When $t_{\mathrm{diff}}\lesssim P_{\mathrm{QPE}}$, where $P_{\mathrm{QPE}}$ is the observed QPE recurrence period, the UV responses from successive disk crossings remain sufficiently separated to preserve a clear cycle-by-cycle correspondence with the X-ray QPEs. When $t_{\mathrm{diff}}\gtrsim P_{\mathrm{QPE}}$, each crossing still produces a UV response, but the individual responses are broadened and overlap with those from subsequent crossings. The resulting UV emission then shows a reduced periodic modulation and appears as a slowly varying or nearly steady component.

\section{Discussion and Conclusions}
\label{sec:discussion}

Our two-channel sBH--disk-transit model connects the soft X-ray and delayed UV components to two distinct physical processes triggered by the same disk crossing. Gravitational focusing and dynamical drag transfer orbital energy to the disk gas and generate hot, optically thick, expanding ejecta. Diffusion from the expanding ejecta produces the soft X-ray QPE in the model. For the fiducial Bondi radius adopted here, we obtain $t_{\mathrm{ej}}\simeq2.1\times10^3\ \mathrm{s}$ and $L_{\mathrm X}\simeq9.6\times10^{41}\ \mathrm{erg\,s^{-1}}$. These estimates reproduce the characteristic short duration and soft X-ray luminosity of QPEs. The transverse motion of the sBH also compresses and bends the background magnetic field. When $P_{\mathrm{ram}}\gtrsim P_{\mathrm{B}_0}$, magnetic flux accumulates near current sheets and triggers in-disk reconnection. For the fiducial parameters, the poloidal and toroidal configurations yield $L_{\mathrm{th,pol}}\simeq6.9\times10^{41}\ \mathrm{erg\,s^{-1}}$ and $L_{\mathrm{th,tor}}\simeq2.7\times10^{43}\ \mathrm{erg\,s^{-1}}$, respectively. The larger $v_{\perp B}$ of the toroidal configuration at the adopted inclination produces stronger compression. 

The soft X-ray and delayed UV components therefore arise from distinct regions and physical processes associated with the same disk crossing. The soft X-ray emission is produced primarily by the expanding ejecta above the disk, whereas the UV emission originates from reconnection-heated material inside the disk and is released after photon diffusion. This differs from shock-cooling models in which the same expanding material can radiate predominantly in either X-ray or the UV under different conditions \citep{2024ApJ...963L...1L}. In a single-ejecta interpretation, the X-ray and UV components originate from the same expanding material and may be influenced by the same underlying ejecta evolution \citep{2024ApJ...963L...1L,2026ApJ..1000L..57G}. In our model, the two bands arise from distinct physical channels, allowing greater freedom in their relative amplitudes and timescales. Simultaneous X-ray and UV monitoring over multiple QPE cycles can therefore help distinguish the two scenarios.

For an Ansky-specific benchmark, we adopt \(P_{\rm QPE}\simeq12\) days measured during the 2025 June--August UV-counterpart campaign \citep{2026ApJ..1001L...6C}. Assuming that the two disk crossings of an inclined circular orbit both produce observable eruptions \citep{2023ApJ...957...34L,2026ApJ..1006L..50L}, the orbital period is $P_{\rm orb}\simeq24$ days. From Kepler's law, $R=\left[GM_\bullet\left(P_{\rm orb}/{2\pi}\right)^2\right]^{1/3} \simeq2.8\times10^2R_{\rm s}$ for $M_\bullet=5\times10^6M_\odot$. Within the adopted reference disk parameters, reproducing the observed X-ray duration and energetics motivates a secondary mass larger than the $10^3M_\odot$ reference normalization. We therefore adopt $m_{\rm sBH}=10^4M_\odot$ as an illustrative Ansky benchmark. Keeping the reference inclination $i=18^\circ$ gives $t_{\rm ej}\simeq1.3$ days, comparable to the observed X-ray flare duration. Using the alternative inner cutoff $b_{\min}\simeq Gm_{\rm sBH}/c^2$ discussed above, the corresponding X-ray energy and characteristic luminosity are $10^{48}\ {\rm erg}$ and $\sim9\times10^{42}\ {\rm erg\,s^{-1}}$, respectively. For the same parameters, the reconnection heating rates are $L_{\rm th,pol}\sim10^{41}\,{\rm erg\,s^{-1}}$ and $L_{\rm th,tor}\sim4\times10^{42}\,{\rm erg\,s^{-1}}$, with the latter providing an energy budget for the variable UV component. The approximately one-day UV lag in Ansky may reflect delayed photon release following in-disk heating \citep{2026ApJ..1000L..57G}. For the same benchmark parameters, the full photon-diffusion time from the disk midplane to the surface is $t_{\rm diff}\simeq7.9\times10^5~{\rm s}$, which may be regarded as an upper-limit estimate for deep energy deposition. To leading order, the observed lag can be interpreted as the difference between the characteristic release times of the two emission channels. The sBH transit may reduce the overlying column density, while reconnection above the disk midplane may shorten the diffusion path. The effective UV diffusion time could therefore be shorter than the full midplane-to-surface estimate, helping to account for the observed short lag.

Apart from Ansky, repeated UV brightenings clearly associated with successive X-ray QPEs have not been commonly detected \citep{2024A&A...690A..80A,2025ApJ...989...13A,2024Natur.634..804N,2025PASA...42e.130G}. Tentative UV dips have also been reported around the X-ray eruptions of XMMSL1 J024916.6--041244 \citep{2021ApJ...921L..40C}. In our model, optically thick ejecta may partially obscure the underlying UV-emitting region under a favorable viewing geometry, producing an apparent UV dip. Given the tentative UV variability and the uncertain covering fraction, we regard this only as a qualitative possibility. In GSN 069, a delayed lower-amplitude X-ray excess was observed  $\sim10$ ks after the primary QPEs, with a period consistent with the recurrence timescale and an estimated luminosity of order $10^{41}\ {\rm erg\,s^{-1}}$ \citep{2023A&A...670A..93M,2023A&A...674L...1M}, which is compatible with the reconnection-heating power in our model. Reconnection occurring in disk surface layers or low-density coronal regions may produce extreme UV or soft X-ray emission \citep{2003ApJ...587..571L,2013ApJ...773...23Z}. The delayed X-ray excess in GSN 069 may therefore arise from the same secondary magnetic-dissipation channel under lower-optical-depth conditions.

Within the present model, a weak background field reduces the reconnection power and allows the UV variability to be diluted by the quiescent nuclear continuum. The diffusion time instead controls the temporal morphology: individual responses remain separable when $t_{\mathrm{diff}}\lesssim P_{\mathrm{QPE}}$, whereas they overlap and produce weakly modulated or nearly steady UV emission when $t_{\mathrm{diff}}\gtrsim P_{\mathrm{QPE}}$. These considerations also lead to several testable observational predictions. The detectability of a UV counterpart depends on both the reconnection-heating strength and the condition $t_{\rm diff}\lesssim P_{\rm QPE}$, which allows UV responses from successive disk crossings to remain temporally distinguishable. A higher $\dot m$ enhances reconnection heating but also increases the optical depth and diffusion time, while a lower orbital inclination weakens magnetic compression and reconnection heating. Thus, constraints on $\dot m$, the disk structure, and the disk--orbit geometry, together with high-cadence multi-cycle X-ray/UV monitoring, can help assess the expected UV behavior of QPE sources.

QPE recurrence intervals are not necessarily constant. In recurrent disk-crossing scenarios, gas drag and gravitational-wave emission dissipate orbital energy and lead to a slow decrease in the intrinsic orbital period. By contrast, apsidal and nodal precession, together with changes in the disk orientation, can shift the crossing locations and alter the impact geometry, producing complex or even nonmonotonic variations in the observed recurrence intervals without requiring comparable changes in the intrinsic orbital period \citep{2024A&A...690A..80A,2025ApJ...985..242Z}. The positive period derivative observed in Ansky has the opposite sign to the orbital decay expected from purely dissipative processes \citep{2026ApJ..1001L...6C}. Our model accounts for its flare duration and delayed UV counterpart for suitable instantaneous disk--orbit parameters, but does not address its secular period increase, which may require additional evolution of the orbit, disk structure, or disk-crossing geometry.

On longer timescales, repeated disk crossings continuously exchange orbital energy and angular momentum, gradually aligning the sBH orbit with the disk \citep{2026PhRvL...}. Hydrodynamic simulations indicate that, at intermediate inclinations, the fractional inclination loss per crossing increases toward lower inclinations. Once the orbit enters the low-inclination regime, $\sin i\lesssim3h$, partially embedded orbits may undergo approximately exponential inclination damping \citep{2025MNRAS.543..132R,2025MNRAS.543.3768W}. Geometrically, when $|\sin i|\lesssim h$, the sBH orbit remains largely confined within the disk body, and the discrete disk-crossing QPE phase may terminate. A lower inclination does not necessarily strengthen the sBH-induced perturbation of the poloidal field. In our corotating relative-velocity prescription, $v_{\perp B,\mathrm{pol}}$ decreases with decreasing inclination, so the instantaneous magnetic compression and reconnection heating are expected to weaken. The longer residence time within the disk may instead make the magnetic perturbation more continuous, but the net magnetic-energy dissipation and radiative output in the embedded phase cannot be inferred from the present local disk-crossing prescription and require a global magnetohydrodynamic (MHD) treatment.

Moreover, the sBH--SMBH system considered here is a persistent source of low-frequency gravitational waves \citep{2026PASJ...78..185S,2026ApJ..1004..221L}. For a nearly circular orbit, the dominant source-frame frequency is $f_{\mathrm{GW}}=\frac{1}{\pi}
\left(\frac{GM_\bullet}{R^3}\right)^{1/2} \simeq 6.5\times10^{-3} M_7^{-1}r_{50}^{-3/2}\ \mathrm{mHz}.$ For the fiducial parameters, this frequency lies below the most sensitive millihertz band of space-based gravitational-wave detectors. Smaller orbital radii or higher harmonics generated by eccentric motion can shift part of the radiated power to higher frequencies and thereby improve the prospects for detection.

Our treatment adopts parameterized orbital, disk, and magnetic-field structures together with a phenomenological multiple-current-sheet prescription. It does not self-consistently follow the long-term orbital alignment, disk capture, or the magnetic and radiative response of the embedded phase. Quantitative tests will require global radiation-MHD simulations coupled to orbital evolution and frequency-dependent radiative transfer calculations.

In summary, we present a model in which a single sBH transit through a magnetized galactic nucleus accretion disk produces two radiative channels. Diffusion from expanding ejecta powers the soft X-ray QPE, while, for a suitable range of background magnetic-field strengths, in-disk magnetic reconnection and photon diffusion produce a delayed UV component. The two channels share the same disk-crossing clock but have different amplitudes and radiative response times. Photon diffusion delays and broadens the UV response relative to the corresponding X-ray QPE. When the UV responses from successive crossings remain resolvable, a clear cycle-by-cycle correspondence between the X-ray and UV emission can be preserved. Longer diffusion times cause adjacent UV responses to overlap, thereby weakening the periodic UV modulation. For an Ansky-specific benchmark, the model reproduces the observed day-scale X-ray duration and energetics, while the reconnection channel provides the energy budget and delay required for the variable UV component. In the future, we will attempt to discuss the complex period evolution from the perspective of accretion disk warping induced by the Zeipel-Lidov-Kozai effect in SMBH binary systems.

\begin{acknowledgments}
This work was supported by the National Key R\&D Program of China (Grant No. 2023YFA1607902), and the National Natural Science Foundation of China (Grant Nos. 12494572, 12221003, 12273089, 12373070, and 12192223).
\end{acknowledgments}


\begin{thebibliography}{}
\expandafter\ifx\csname natexlab\endcsname\relax\def\natexlab#1{#1}\fi
\providecommand{\url}[1]{\href{#1}{#1}}
\providecommand{\dodoi}[1]{doi:~\href{http://doi.org/#1}{\nolinkurl{#1}}}
\providecommand{\doeprint}[1]{\href{http://ascl.net/#1}{\nolinkurl{http://ascl.net/#1}}}
\providecommand{\doarXiv}[1]{\href{https://arxiv.org/abs/#1}{\nolinkurl{https://arxiv.org/abs/#1}}}

\bibitem[{R. {Arcodia} {et~al.}(2025){Arcodia}, {Baldini}, {Merloni}, {et~al.}}]{2025ApJ...989...13A} {Arcodia}, R., {Baldini}, P., {Merloni}, A., {et~al.} 2025, \bibinfo{title}{{SRG/eROSITA No. 5: Discovery of Quasiperiodic Eruptions Every $\sim$3.7 Days from a Galaxy at $z>0.1$},} \apj, 989, 13, \dodoi{10.3847/1538-4357/adec9b}

\bibitem[{R. {Arcodia} {et~al.}(2024{\natexlab{a}}){Arcodia}, {Linial}, {Miniutti}, {et~al.}}]{2024A&A...690A..80A} {Arcodia}, R., {Linial}, I., {Miniutti}, G., {et~al.} 2024{\natexlab{a}}, \bibinfo{title}{{Ticking Away: The Long-term X-Ray Timing and Spectral Evolution of eRO-QPE2},} \aap, 690, A80, \dodoi{10.1051/0004-6361/202451218}

\bibitem[{R. {Arcodia} {et~al.}(2024{\natexlab{b}}){Arcodia}, {Liu}, {Merloni}, {et~al.}}]{2024A&A...684A..64A} {Arcodia}, R., {Liu}, Z., {Merloni}, A., {et~al.} 2024{\natexlab{b}}, \bibinfo{title}{{The More the Merrier: SRG/eROSITA Discovers Two Further Galaxies Showing X-Ray Quasi-periodic Eruptions},} \aap, 684, A64, \dodoi{10.1051/0004-6361/202348881}

\bibitem[{R. {Arcodia} {et~al.}(2021){Arcodia}, {Merloni}, {Nandra}, {et~al.}}]{2021Natur.592..704A} {Arcodia}, R., {Merloni}, A., {Nandra}, K., {et~al.} 2021, \bibinfo{title}{{X-ray Quasi-periodic Eruptions from Two Previously Quiescent Galaxies},} \nat, 592, 704, \dodoi{10.1038/s41586-021-03394-6}

\bibitem[{R. {Arcodia} {et~al.}(2022){Arcodia}, {Miniutti}, {Ponti}, {et~al.}}]{2022A&A...662A..49A} {Arcodia}, R., {Miniutti}, G., {Ponti}, G., {et~al.} 2022, \bibinfo{title}{{The Complex Time and Energy Evolution of Quasi-periodic Eruptions in eRO-QPE1},} \aap, 662, A49, \dodoi{10.1051/0004-6361/202243259}

\bibitem[{W.~D. {Arnett}(1982{\natexlab{a}}){Arnett}}]{1982ApJ...253..785A} {Arnett}, W.~D. 1982{\natexlab{a}}, \bibinfo{title}{{Type I Supernovae. I. Analytic Solutions for the Early Part of the Light Curve},} \apj, 253, 785, \dodoi{10.1086/159681}

\bibitem[{P. {Baldini} {et~al.}(2026){Baldini}, {Rau}, {Merloni}, {et~al.}}]{2026A&A...706L..15B} {Baldini}, P., {Rau}, A., {Merloni}, A., {et~al.} 2026, \bibinfo{title}{{Discovery of Crested Quasi-periodic Eruptions Following the Most Luminous SRG/eROSITA Tidal Disruption Event},} \aap, 706, L15, \dodoi{10.1051/0004-6361/202558241}

\bibitem[{H. {Bondi}(1952){Bondi}}]{1952MNRAS.112..195B} {Bondi}, H. 1952, \bibinfo{title}{{On Spherically Symmetrical Accretion},} \mnras, 112, 195, \dodoi{10.1093/mnras/112.2.195}

\bibitem[{H. {Bondi} {\&} F. {Hoyle}(1944){Bondi} \& {Hoyle}}]{1944MNRAS.104..273B} {Bondi}, H., \& {Hoyle}, F. 1944, \bibinfo{title}{{On the Mechanism of Accretion by Stars},} \mnras, 104, 273, \dodoi{10.1093/mnras/104.5.273}

\bibitem[{J. {Chakraborty} {et~al.}(2024){Chakraborty}, {Arcodia}, {Kara}, {et~al.}}]{2024ApJ...965...12C} {Chakraborty}, J., {Arcodia}, R., {Kara}, E., {et~al.} 2024, \bibinfo{title}{{Testing EMRI Models for Quasi-periodic Eruptions with 3.5 yr of Monitoring eRO-QPE1},} \apj, 965, 12, \dodoi{10.3847/1538-4357/ad2941}

\bibitem[{J. {Chakraborty} {et~al.}(2025){Chakraborty}, {Kara}, {Arcodia}, {et~al.}}]{2025ApJ...983L..39C} {Chakraborty}, J., {Kara}, E., {Arcodia}, R., {et~al.} 2025, \bibinfo{title}{{Discovery of Quasiperiodic Eruptions in the Tidal Disruption Event and Extreme Coronal Line Emitter AT2022upj: Implications for the QPE/TDE Fraction and a Connection to ECLEs},} \apjl, 983, L39, \dodoi{10.3847/2041-8213/adc2f8}

\bibitem[{J. {Chakraborty} {et~al.}(2021){Chakraborty}, {Kara}, {Masterson}, {et~al.}}]{2021ApJ...921L..40C} {Chakraborty}, J., {Kara}, E., {Masterson}, M., {et~al.} 2021, \bibinfo{title}{{Possible Quasi-periodic Eruptions in a Tidal Disruption Event Candidate},} \apjl, 921, L40, \dodoi{10.3847/2041-8213/ac313b}

\bibitem[{J. {Chakraborty} {et~al.}(2026){Chakraborty}, {Rappaport}, {Arcodia}, {et~al.}}]{2026ApJ..1001L...6C} {Chakraborty}, J., {Rappaport}, S.~A., {Arcodia}, R., {et~al.} 2026, \bibinfo{title}{{A Positive Period Derivative in the Quasiperiodic Eruptions of ZTF19acnskyy},} \apjl, 1001, L6, \dodoi{10.3847/2041-8213/ae548b}

\bibitem[{L.~J. {Dursi} {\&} C. {Pfrommer}(2008){Dursi} \& {Pfrommer}}]{2008ApJ...677..993D} {Dursi}, L.~J., \& {Pfrommer}, C. 2008, \bibinfo{title}{{Draping of Cluster Magnetic Fields over Bullets and Bubbles: Morphology and Dynamic Effects},} \apj, 677, 993, \dodoi{10.1086/529371}

\bibitem[{I. {El Mellah} {et~al.}(2023){El Mellah}, {Cerutti}, \& {Crinquand}}]{2023A&A...677A..67E} {El Mellah}, I., {Cerutti}, B., \& {Crinquand}, B. 2023, \bibinfo{title}{{Reconnection-driven Flares in 3D Black Hole Magnetospheres: A Scenario for Hot Spots around Sagittarius A*},} \aap, 677, A67, \dodoi{10.1051/0004-6361/202346781}

\bibitem[{I. {El Mellah} {et~al.}(2022){El Mellah}, {Cerutti}, {Crinquand}, \& {Parfrey}}]{2022A&A...663A.169E} {El Mellah}, I., {Cerutti}, B., {Crinquand}, B., \& {Parfrey}, K. 2022, \bibinfo{title}{{Spinning Black Holes Magnetically Connected to a Keplerian Disk: Magnetosphere, Reconnection Sheet, Particle Acceleration, and Coronal Heating},} \aap, 663, A169, \dodoi{10.1051/0004-6361/202142847}

\bibitem[{A. {Franchini} {et~al.}(2023){Franchini}, {Bonetti}, {Lupi}, {et~al.}}]{2023A&A...675A.100F} {Franchini}, A., {Bonetti}, M., {Lupi}, A., {et~al.} 2023, \bibinfo{title}{{Quasi-periodic Eruptions from Impacts between the Secondary and a Rigidly Precessing Accretion Disc in an Extreme Mass-ratio Inspiral System},} \aap, 675, A100, \dodoi{10.1051/0004-6361/202346565}

\bibitem[{P. {Ghosh} {\&} M.~A. {Abramowicz}(1997){Ghosh} \& {Abramowicz}}]{1997MNRAS.292..887G} {Ghosh}, P., \& {Abramowicz}, M.~A. 1997, \bibinfo{title}{{Electromagnetic Extraction of Rotational Energy from Disc-fed Black Holes: The Strength of the Blandford--Znajek Process},} \mnras, 292, 887, \dodoi{10.1093/mnras/292.4.887}

\bibitem[{M. {Giustini} {et~al.}(2024){Giustini}, {Miniutti}, {Arcodia}, {et~al.}}]{2024A&A...692A..15G} {Giustini}, M., {Miniutti}, G., {Arcodia}, R., {et~al.} 2024, \bibinfo{title}{{Fragments of Harmony amid Apparent Chaos: A Closer Look at the X-Ray Quasi-periodic Eruptions of the Galaxy RX J1301.9+2747},} \aap, 692, A15, \dodoi{10.1051/0004-6361/202450861}

\bibitem[{M. {Giustini} {et~al.}(2020){Giustini}, {Miniutti}, \& {Saxton}}]{2020A&A...636L...2G} {Giustini}, M., {Miniutti}, G., \& {Saxton}, R.~D. 2020, \bibinfo{title}{{X-ray Quasi-periodic Eruptions from the Galactic Nucleus of RX J1301.9+2747},} \aap, 636, L2, \dodoi{10.1051/0004-6361/202037610}



\bibitem[{A.~J. {Goodwin} {et~al.}(2025){Goodwin}, {Arcodia}, {Miniutti}, {Miller-Jones}, \& {van Velzen}}]{2025PASA...42e.130G} {Goodwin}, A.~J., {Arcodia}, R., {Miniutti}, G., {Miller-Jones}, J.~C.~A., \& {van Velzen}, S. 2025, \bibinfo{title}{{The Radio Properties of Quasi-periodic X-Ray Eruption Sources},} \pasa, 42, e130, \dodoi{10.1017/pasa.2025.10083}

\bibitem[{H. {Guo} {et~al.}(2026){Guo}, {Yan}, {Li}, {et~al.}}]{2026ApJ..1000L..57G} {Guo}, H., {Yan}, Z., {Li}, Y.-P., {et~al.} 2026, \bibinfo{title}{{Evidence for a Delayed Ultraviolet Counterpart to X-Ray Quasiperiodic Eruptions in Ansky},} \apjl, 1000, L57, \dodoi{10.3847/2041-8213/ae524b}

\bibitem[{L. {Hern{\'a}ndez-Garc{\'i}a} {et~al.}(2025{\natexlab{a}}){Hern{\'a}ndez-Garc{\'i}a}, {Chakraborty}, {S{\'a}nchez-S{\'a}ez}, {et~al.}}]{2025NatAs...9..895H} {Hern{\'a}ndez-Garc{\'i}a}, L., {Chakraborty}, J., {S{\'a}nchez-S{\'a}ez}, P., {et~al.} 2025{\natexlab{a}}, \bibinfo{title}{{Discovery of Extreme Quasi-periodic Eruptions in a Newly Accreting Massive Black Hole},} Nature Astronomy, 9, 895, \dodoi{10.1038/s41550-025-02523-9}

\bibitem[{L. {Hern{\'a}ndez-Garc{\'i}a} {et~al.}(2025{\natexlab{b}}){Hern{\'a}ndez-Garc{\'i}a}, {S{\'a}nchez-S{\'a}ez}, {Chakraborty}, {et~al.}}]{2025A&A...703A.263H} {Hern{\'a}ndez-Garc{\'i}a}, L., {S{\'a}nchez-S{\'a}ez}, P., {Chakraborty}, J., {et~al.} 2025{\natexlab{b}}, \bibinfo{title}{{NICER Observations Reveal Doubled Timescales in Ansky's Quasi-periodic Eruptions},} \aap, 703, A263, \dodoi{10.1051/0004-6361/202555258}

\bibitem[{B.~F. {Liu} {et~al.}(2003){Liu}, {Mineshige}, \& {Ohsuga}}]{2003ApJ...587..571L}{Liu}, B.~F., {Mineshige}, S., \& {Ohsuga}, K. 2003,\bibinfo{title}{{Spectra from a Magnetic Reconnection-Heated Corona in Active Galactic Nuclei},}The Astrophysical Journal, 587, 571,
\dodoi{10.1086/368282}

\bibitem[{L. {Lui} {et~al.}(2026){Lui}, {Torres-Orjuela}, {Kar Chowdhury}, \& {Dai}}]{2026ApJ..1004..221L} {Lui}, L., {Torres-Orjuela}, A., {Kar Chowdhury}, R., \& {Dai}, L. 2026,\bibinfo{title}{{Gravitational-wave Signatures of Quasiperiodic Eruptions: LISA Detection Prospects for RX J1301.9+2747},}\apj, 1004, 221,
\dodoi{10.3847/1538-4357/ae71cc}

\bibitem[{L.{Lui}{et~al.}(2026){Lui},{Drummond},\&{Torres-Orjuela}}]{2026PhRvL...}{Lui}, L., {Drummond}, L.~V., \& {Torres-Orjuela}, A. 2026,\bibinfo{title}{{Pitching Cosmic Curveballs: Environmental Effects on Extreme-Mass-Ratio Inspirals with Spinning Secondaries},}\prl, accepted,
\dodoi{10.1103/dy3h-65ty}

\bibitem[{H.~K. {Lee} {et~al.}(1999){Lee}, {Wijers}, \& {Brown}}]{1999ASPC..190..173L} {Lee}, H.~K., {Wijers}, R.~A.~M.~J., \& {Brown}, G.~E. 1999, \bibinfo{title}{{The Blandford--Znajek Process as a Gamma-Ray Burst Central Engine},} in Astronomical Society of the Pacific Conference Series, Vol. 190, Gamma-Ray Bursts: The First Three Minutes, ed. J.~{Poutanen} \& R.~{Svensson} (San Francisco, CA: Astronomical Society of the Pacific), 173

\bibitem[{I. {Linial} {\&} B.~D. {Metzger}(2023){Linial} \& {Metzger}}]{2023ApJ...957...34L} {Linial}, I., \& {Metzger}, B.~D. 2023, \bibinfo{title}{{EMRI + TDE = QPE: Periodic X-Ray Flares from Star--Disk Collisions in Galactic Nuclei},} \apj, 957, 34, \dodoi{10.3847/1538-4357/acf65b}

\bibitem[{I. {Linial} {\&} B.~D. {Metzger}(2024){Linial} \& {Metzger}}]{2024ApJ...963L...1L} {Linial}, I., \& {Metzger}, B.~D. 2024, \bibinfo{title}{{Ultraviolet Quasiperiodic Eruptions from Star--Disk Collisions in Galactic Nuclei},} \apjl, 963, L1, \dodoi{10.3847/2041-8213/ad2464}

\bibitem[{K. {Liu} {et~al.}(2026){Liu}, {Liu}, {Pan}, {Deng}, {Shen}, \& {Yu}}]{2026ApJ..1006L..50L} {Liu}, K., {Liu}, S.-F., {Pan}, Z., {et~al.} 2026, \bibinfo{title}{{Quasiperiodic Eruptions from Stellar-mass Black Holes Impacting Accretion Disks in Galactic Nuclei},} \apjl, 1006, L50, \dodoi{10.3847/2041-8213/ae8aeb}

\bibitem[{T. {Liu} {et~al.}(2017){Liu}, {Gu}, \& {Zhang}}]{2017NewAR..79....1L} {Liu}, T., {Gu}, W.-M., \& {Zhang}, B. 2017, \bibinfo{title}{{Neutrino-dominated Accretion Flows as the Central Engine of Gamma-Ray Bursts},} \nar, 79, 1, \dodoi{10.1016/j.newar.2017.07.001}


\bibitem[{G. {Miniutti} {et~al.}(2019){Miniutti}, {Saxton}, {Giustini}, {et~al.}}]{2019Natur.573..381M} {Miniutti}, G., {Saxton}, R.~D., {Giustini}, M., {et~al.} 2019, \bibinfo{title}{{Nine-hour X-ray Quasi-periodic Eruptions from a Low-mass Black Hole Galactic Nucleus},} \nat, 573, 381, \dodoi{10.1038/s41586-019-1556-x}


\bibitem[{G. {Miniutti} {et~al.}(2023a){Miniutti}, {Giustini}, {Arcodia}, {et~al.}}]{2023A&A...670A..93M}{Miniutti}, G., {Giustini}, M., {Arcodia}, R., {Saxton}, R.~D.,{Read}, A.~M., {Bianchi}, S., \& {Alexander}, K.~D. 2023a,\bibinfo{title}{{Repeating tidal disruptions in GSN 069: Long-term evolution and constraints on quasi-periodic eruptions' models},} Astronomy \& Astrophysics, 670, A93,
\dodoi{10.1051/0004-6361/202244512}



\bibitem[{G. {Miniutti} {et~al.}(2023b){Miniutti}, {Giustini}, {Arcodia}, {et~al.}}]{2023A&A...674L...1M} {Miniutti}, G., {Giustini}, M., {Arcodia}, R., {et~al.} 2023, \bibinfo{title}{{Alive and Kicking: A New QPE Phase in GSN 069 Revealing a Quiescent Luminosity Threshold for QPEs},} \aap, 674, L1, \dodoi{10.1051/0004-6361/202346653}

\bibitem[{A. {Nathanail} {et~al.}(2022){Nathanail}, {Mpisketzis}, {Porth}, {Fromm}, \& {Rezzolla}}]{2022MNRAS.513.4267N} {Nathanail}, A., {Mpisketzis}, V., {Porth}, O., {Fromm}, C.~M., \& {Rezzolla}, L. 2022, \bibinfo{title}{{Magnetic Reconnection and Plasmoid Formation in Three-dimensional Accretion Flows around Black Holes},} \mnras, 513, 4267, \dodoi{10.1093/mnras/stac1118}

\bibitem[{L. {Ni} {et~al.}(2018){Ni}, {Lukin}, {Murphy}, \& {Lin}}]{2018ApJ...852...95N} {Ni}, L., {Lukin}, V.~S., {Murphy}, N.~A., \& {Lin}, J. 2018, \bibinfo{title}{{Magnetic Reconnection in Strongly Magnetized Regions of the Low Solar Chromosphere},} \apj, 852, 95, \dodoi{10.3847/1538-4357/aa9edb}

\bibitem[{M. {Nicholl} {et~al.}(2024){Nicholl}, {Pasham}, {Mummery}, {et~al.}}]{2024Natur.634..804N} {Nicholl}, M., {Pasham}, D.~R., {Mummery}, A., {et~al.} 2024, \bibinfo{title}{{Quasi-periodic X-Ray Eruptions Years after a Nearby Tidal Disruption Event},} \nat, 634, 804, \dodoi{10.1038/s41586-024-08023-6}

\bibitem[{X. {Pan} {et~al.}(2023){Pan}, {Li}, \& {Cao}}]{2023ApJ...952...32P} {Pan}, X., {Li}, S.-L., \& {Cao}, X. 2023, \bibinfo{title}{{Application of the Disk Instability Model to All Quasiperiodic Eruptions},} \apj, 952, 32, \dodoi{10.3847/1538-4357/acd180}
    
\bibitem[{X. {Pan} {et~al.}(2022){Pan}, {Li}, {Cao}, {Miniutti}, \& {Gu}}]{2022ApJ...928L..18P} {Pan}, X., {Li}, S.-L., {Cao}, X., {et~al.} 2022, \bibinfo{title}{{A Disk Instability Model for the Quasi-periodic Eruptions of GSN 069},} \apjl, 928, L18, \dodoi{10.3847/2041-8213/ac5faf}

\bibitem[{X. {Pan} {et~al.}(2025){Pan}, {Li}, {Cao}, {Liu}, \& {Yuan}}]{2025ApJ...989..196P} {Pan}, X., {Li}, S.-L., {Cao}, X., {et~al.} 2025, \bibinfo{title}{{Disk Instability Model for Quasi-periodic Eruptions: Investigating Period Dispersion and Peak Temperature},} \apj, 989, 196, \dodoi{10.3847/1538-4357/adf05d}

\bibitem[{B. {Ripperda} {et~al.}(2020){Ripperda}, {Bacchini}, \& {Philippov}}]{2020ApJ...900..100R} {Ripperda}, B., {Bacchini}, F., \& {Philippov}, A.~A. 2020, \bibinfo{title}{{Magnetic Reconnection and Hot Spot Formation in Black Hole Accretion Disks},} \apj, 900, 100, \dodoi{10.3847/1538-4357/ababab}

\bibitem[{B. {Ripperda} {et~al.}(2022){Ripperda}, {Liska}, {Chatterjee}, {et~al.}}]{2022ApJ...924L..32R} {Ripperda}, B., {Liska}, M., {Chatterjee}, K., {et~al.} 2022, \bibinfo{title}{{Black Hole Flares: Ejection of Accreted Magnetic Flux through 3D Plasmoid-mediated Reconnection},} \apjl, 924, L32, \dodoi{10.3847/2041-8213/ac46a1}

\bibitem[{B. {Ripperda} {et~al.}(2021){Ripperda}, {Mahlmann}, {Chernoglazov}, {et~al.}}]{2021JPlPh..87e9012R} {Ripperda}, B., {Mahlmann}, J.~F., {Chernoglazov}, A., {et~al.} 2021, \bibinfo{title}{{Weak Alfv{\'e}nic Turbulence in Relativistic Plasmas. Part 2. Current Sheets and Dissipation},} Journal of Plasma Physics, 87, 905870512, \dodoi{10.1017/S0022377821000957}

\bibitem[{C. {Rowan} {et~al.}(2025){Rowan}, {Whitehead}, {Fabj}, {Kirkeberg}, {Pessah}, \& {Kocsis}}]{2025MNRAS.543..132R} {Rowan}, C., {Whitehead}, H., {Fabj}, G., {et~al.} 2025, \bibinfo{title}{{Hydrodynamic Simulations of Black Hole Evolution in AGN Discs. I. Orbital Alignment of Highly Inclined Satellites},} \mnras, 543, 132, \dodoi{10.1093/mnras/staf1449}

\bibitem[{G.~B. {Rybicki} {\&} A.~P. {Lightman}(1979){Rybicki} \& {Lightman}}]{1979rpa..book.....R} {Rybicki}, G.~B., \& {Lightman}, A.~P. 1979, {Radiative Processes in Astrophysics} (New York: Wiley-Interscience)

\bibitem[{P. {S{\'a}nchez-S{\'a}ez} {et~al.}(2024){S{\'a}nchez-S{\'a}ez}, {Hern{\'a}ndez-Garc{\'i}a}, {Bernal}, {et~al.}}]{2024A&A...688A.157S} {S{\'a}nchez-S{\'a}ez}, P., {Hern{\'a}ndez-Garc{\'i}a}, L., {Bernal}, S., {et~al.} 2024, \bibinfo{title}{{SDSS1335+0728: The Awakening of a $\sim 10^6\,M_\odot$ Black Hole},} \aap, 688, A157, \dodoi{10.1051/0004-6361/202347957}

\bibitem[{N.~I. {Shakura} {\&} R.~A. {Sunyaev}(1973){Shakura} \& {Sunyaev}}]{1973A&A....24..337S} {Shakura}, N.~I., \& {Sunyaev}, R.~A. 1973, \bibinfo{title}{{Black Holes in Binary Systems. Observational Appearance},} \aap, 24, 337

\bibitem[{T. {Suzuguchi} {et~al.}(2026){Suzuguchi}, {Omiya}, \& {Takeda}}]{2026PASJ...78..185S} {Suzuguchi}, T., {Omiya}, H., \& {Takeda}, H. 2026, \bibinfo{title}{{Possibility of Multi-messenger Observations of Quasi-periodic Eruptions with X-Rays and Gravitational Waves},} \pasj, 78, 185, \dodoi{10.1093/pasj/psaf132}

\bibitem[{D. {Thun} {et~al.}(2016){Thun}, {Kuiper}, {Schmidt}, \& {Kley}}]{2016A&A...589A..10T} {Thun}, D., {Kuiper}, R., {Schmidt}, F., \& {Kley}, W. 2016, \bibinfo{title}{{Dynamical Friction for Supersonic Motion in a Homogeneous Gaseous Medium},} \aap, 589, A10, \dodoi{10.1051/0004-6361/201527629}

\bibitem[{I. {Vurm} {et~al.}(2025){Vurm}, {Linial}, \& {Metzger}}]{2025ApJ...983...40V} {Vurm}, I., {Linial}, I., \& {Metzger}, B.~D. 2025, \bibinfo{title}{{Radiation Transport Simulations of Quasiperiodic Eruptions from Star--Disk Collisions},} \apj, 983, 40, \dodoi{10.3847/1538-4357/adb74d}

\bibitem[{H. {Whitehead} {et~al.}(2025){Whitehead}, {Rowan}, \& {Kocsis}}]{2025MNRAS.543.3768W} {Whitehead}, H., {Rowan}, C., \& {Kocsis}, B. 2025, \bibinfo{title}{{Hydrodynamic Simulations of Black Hole Evolution in AGN Discs. II. Inclination Damping for Partially Embedded Satellites},} \mnras, 543, 3768, \dodoi{10.1093/mnras/staf1686}

\bibitem[{J.-T. {Xing} {et~al.}(2025){Xing}, {Liu}, {Huang}, \& {Sun}}]{2025ApJ...991..167X} {Xing}, J.-T., {Liu}, T., {Huang}, B.-Q., \& {Sun}, M. 2025, \bibinfo{title}{{Variabilities Driven by Satellite Black Hole Migration in Active Galactic Nucleus Disks},} \apj, 991, 167, \dodoi{10.3847/1538-4357/adff71}

\bibitem[{X. {Zhong} \& J. {Wang}(2013){Zhong}\&{Wang}}]{2013ApJ...773...23Z}{Zhong}, X., \& {Wang}, J. 2013,\bibinfo{title}{{A Magnetic Reconnection Origin for the Soft X-Ray Excess in an Active Galactic Nucleus},}The Astrophysical Journal, 773, 23,
\dodoi{10.1088/0004-637X/773/1/23}

\bibitem[{C. {Zhou} {et~al.}(2024){Zhou}, {Huang}, {Guo}, {Li}, \& {Pan}}]{2024PhRvD.109j3031Z} {Zhou}, C., {Huang}, L., {Guo}, K., {Li}, Y.-P., \& {Pan}, Z. 2024, \bibinfo{title}{{Probing Orbits of Stellar Mass Objects Deep in Galactic Nuclei with Quasi-periodic Eruptions},} \prd, 109, 103031, \dodoi{10.1103/PhysRevD.109.103031}

\bibitem[{C. {Zhou} {et~al.}(2025){Zhou}, {Zeng}, \& {Pan}}]{2025ApJ...985..242Z} {Zhou}, C., {Zeng}, Y., \& {Pan}, Z. 2025, \bibinfo{title}{{Secular Evolution of Quasiperiodic Eruptions},} \apj, 985, 242, \dodoi{10.3847/1538-4357/adcee2}

\end{thebibliography}
\end{document}